# Tunable charge-trap memory based on few-layer MoS$_2$

Enze Zhang[1], Weiyi Wang[1], Cheng Zhang[1], Yibo Jin[1], Guodong Zhu[2], Qingqing Sun[3], David Wei Zhang[3], Peng Zhou[3†] & Faxian Xiu[1†]

[1]State Key Laboratory of Surface Physics and Department of Physics, Fudan University, Shanghai 200433, China

[2]Department of Materials Science, Fudan University, Shanghai 200433, China

[3]State Key Laboratory of ASIC and System, Department of Microelectronics, Fudan University, Shanghai 200433, China

†Correspondence and requests for materials should be addressed to F. X. and P. Z. (E-mails: faxian@fudan.edu.cn, pengzhou@fudan.edu.cn)

**Abstract**

**Charge-trap memory with high-κ dielectric materials is considered to be a promising candidate for next-generation memory devices. Ultrathin layered two-dimensional (2D) materials like graphene and $MoS_2$ have been receiving much attention because of their novel physical properties and potential applications in electronic devices. Here, we report on a dual-gate charge-trap memory device composed of a few-layer $MoS_2$ channel and a three-dimensional (3D) $Al_2O_3/HfO_2/Al_2O_3$ charge-trap gate stack. Owing to the extraordinary trapping ability of both electrons and holes in $HfO_2$, the $MoS_2$ memory device exhibits an unprecedented memory window exceeding 20 V. More importantly, with a back gate the window size can be effectively tuned from 15.6 to 21 V; the program/erase current ratio can reach up to $10^4$, far beyond Si-based flash memory, which allows for multi-bit information storage. Furthermore, the device shows a high mobility of 170 $cm^2V^{-1}s^{-1}$, a good endurance of hundreds of cycles and a stable retention of ~28% charge loss after 10 years which is drastically lower than ever reported $MoS_2$ flash memory. The combination of 2D materials with traditional high-κ charge-trap gate stacks opens up an exciting field of novel nonvolatile memory devices.**

Atomically thin 2D materials like graphene and MoS$_2$ has been extensively studied recently because of their promising applications in optoelectronics[1, 2], spintronics[3-7], transparent and flexible devices[8-12]. Due to its remarkable properties, such as high carrier mobility and mechanical flexibility, graphene has been incorporated into nonvolatile memory structures serving as a floating gate[13, 14] or a transparent channel[15]. However, owing to its zero band gap[16], the graphene channeled memory devices typically possess a low program/erase current ratio, which significantly hinders its application in nonvolatile memory devices. Unlike graphene, MoS$_2$ has a transition from indirect band gap (1.2 eV) to a direct band gap (1.8 eV) in monolayer[17, 18]. Its field effect transistors[19] show a high mobility of 200 cm$^2$V$^{-1}$s$^{-1}$ with a high on/off ratio approximately 10$^8$. To potentially enhance the program/erase current ratio, attempts were made to replace graphene with MoS$_2$ as a channel material in a ferroelectric memory[20] or as a charge-trap layer in a graphene flash memory[21]. It was demonstrated that the monolayer MoS$_2$ is very sensitive to the presence of charges[14]. However, the relatively small memory window, the degraded mobility, and the insufficient trap capability in those devices require further improvement of the charge-trap stack in the MoS$_2$ memory device.

It is known that high-κ dielectrics can serve as excellent charge-trap layers due to their reduced coupling crosstalk, weak charge leakage and good scalability[22, 23]. Recently, various kinds of high-κ dielectrics were investigated including HfO$_2$, TiO$_2$, HfAlO and so on[24-29], among which the Al$_2$O$_3$/HfO$_2$/Al$_2$O$_3$ gate stack has been widely used because of its high transparency, extraordinary thermal stability and remarkable trap capability[30-33]. Previous studies have also shown that the high-κ dielectrics can greatly enhance the mobility of MoS$_2$ transistors[19, 34]. Thus, we anticipate that the integration of MoS$_2$ with Al$_2$O$_3$/HfO$_2$/Al$_2$O$_3$ can significantly enlarge the memory window and enhance the device performance.

In this letter, we demonstrate a nonvolatile dual-gate charge-trap memory based on MoS$_2$ and high-κ HfO$_2$. As expected, the device shows a substantial memory window originated from the superior trapping capacity of the Al$_2$O$_3$/HfO$_2$/Al$_2$O$_3$ stack.

A back-gate electrode was also used to tune the carrier density of MoS$_2$, giving rise to a systematic shift of the threshold voltage and a gate-tunable memory window, thus enabling multi-bit information storage. The application of the conventional Al$_2$O$_3$/HfO$_2$/Al$_2$O$_3$ gate stack renders a possibility for a massive production of high-performance MoS$_2$-based 2D memory devices.

**Results**

**Device preparation and characterizations.** Few-layer MoS$_2$ is obtained through mechanical exfoliation from bulk MoS$_2$ crystals onto pre-patterned SiO$_2$/Si substrates. The thickness of SiO$_2$ is 270 nm. Two Cr/Au electrodes of 12nm/100nm were deposited via *e*-beam evaporation to form source and drain contacts. Subsequently, the Al$_2$O$_3$/HfO$_2$/Al$_2$O$_3$ gate stack was grown via an atomic layer deposition (ALD) system with a layer thickness of 7nm/8nm/30nm, respectively. More details of device fabrication process can be found in the "methods" section and in the supplementary Figure S1. A typical device architecture is shown in Figure 1a (also see a SEM picture in Figure 1b inset). An applied top gate voltage ($V_{TG}$) modulates the amount of charge stored in the HfO$_2$ charge-trap layer, causing the variation of the conductivity of the MoS$_2$ channel. A back-gate voltage ($V_{BG}$) was applied to the degenerately doped silicon substrate to tune the memory characteristics by systematically shifting the Fermi level of MoS$_2$. To check the gate modulation, *I-V* curves at different top-gate voltage were measured (Figure 1b): the source-drain current ($I_{DS}$) varies linearly with source-drain voltage ($V_{DS}$), indicative of developing Ohmic contacts[35, 36]. The transfer curve ($I_{DS}$-$V_{TG}$) of the device can be obtained by sweeping $V_{TG}$ while keeping the back gate grounded (Figure 1c). With $V_{DS}$ gradually rising from 100 to 600 mV, a maximal on/off ratio higher than 10$^5$ was acquired. The calculated field-effect mobility of the memory device is about 170 cm$^2$V$^{-1}$s$^{-1}$. Such a high mobility could be attributed to the encapsulation of MoS$_2$ in a high-κ dielectric environment which reduces the Coulomb scattering and modifies the phonon dispersion in few-layer MoS$_2$[34, 37]. The MoS$_2$ in our device is measured to be about 3-4 layers by Raman spectroscopy (Figure 1c inset)[38]. To elucidate the effect of interface charges on the device performance, it is necessary to measure the hysteresis behavior of $I_{DS}$ - $V_{BG}$ (Figure 1d). When $V_{BG}$ is swept between -40 and +40 V, the overlap of the forward and backward sweep curves shows a clean interface between $SiO_2$ and the $MoS_2$ channel[39].

The transfer characteristics were further explored to probe the storage capability of the $MoS_2$ memory device. As shown in Figure 2a, with $V_{BG}=0$, $V_{TG}$ was swept from -26 to +26 V and back to -26 V. A huge memory window of 20 V was observed, primarily originated from a large amount of electrons and holes stored in the $HfO_2$ charge-trap layer. Figure 2b illuminates the device operation process. When $V_{TG}$ is swept towards a high positive value, electrons can tunnel through the 7 nm-thick $Al_2O_3$ barrier by means of Fowler-Nordheim tunneling[40]. The resultant accumulation of electrons in $HfO_2$ screens the top-gate electric field to reach the $MoS_2$ channel, which results in a positive shift of the threshold voltage (red curve in Figure 2a). When $V_{TG}$ is swept from the positive to the negative direction, however, electrons are transferred back from the charge layer to the channel; simultaneously holes tunnel through the barrier and are trapped in the $HfO_2$ layer which causes the threshold shifting to the negative direction. The capability of electron and hole trapping leads to the appreciable memory window as large as 20 V, which is different from previously reported memory devices using monolayer $MoS_2$ as the channel[14, 21].

The amount of charge stored in the charge-trap layer can be modulated by gradually changing the maximum (+$V_{TG.max}$) and minimum (-$V_{TG.max}$) voltage applied on the top gate. Figure 2c shows the enlarged hysteresis window when |$V_{TG.max}$| becomes larger. As aforementioned, the shift of the threshold voltage towards negative and positive direction corresponds to the hole and the electron trapping, respectively. The deduced threshold voltage shift ($\Delta V$) as a function of $V_{TG.max}$ is summarized in Figure 2d. The amount of charge stored in the charge−trap layer can be estimated from the expression[41] $n = (\Delta V \times C_{HF-AL})/e$, where $e$ is the electron charge, $\Delta V$ is the threshold voltage shift towards the negative or the positive direction compare to the original transfer curve. According to Figure 2c-d, the hysteresis window of the sweep between -10 and +10 V is close to zero, thus it can be defined as the original transfer curve, corresponding to no tunneled electrons or holes existing in the charge-trap layer. $C_{HF-AL} = \varepsilon_0 \varepsilon_{AL}/d_{AL}$ is the capacitance between the HfO$_2$ charge-trap layer and the top gate, where $\varepsilon_0$ is the vacuum permittivity, $\varepsilon_{AL}$ and $d_{AL}$ are the relative dielectric constant (~8) and thickness (~30 nm) of the Al$_2$O$_3$ blocking layer, respectively. The calculated density of stored electrons and holes is on the order of ~8.6×10$^{13}$ cm$^{-2}$ and ~1.9×10$^{14}$ cm$^{-2}$, respectively, which is much higher than previously reported memory devices using graphene or graphene oxide as charge-trap layers.[13, 42] The lower tunneling barrier height of holes than electrons is accounted for the high trap density of holes[43], as detailed in supplementary Figure S2.

To study the dynamic transition rate of the MoS$_2$ memory device[15, 21], a negative pulse (-26 V, duration of 3 s) was applied to the top gate ($V_{BG}$ = 0) to set the device in the erase state, followed by a +26 V pulse with different duration time. The reading procedure was performed by sweeping $I_{DS}$-$V_{TG}$ in a very small range (-5 to +9 V) to minimize the effect of the measurements to the device's state. After each reading operation, a negative pulse (-26 V, duration of 3 s) was applied on the top gate to the reset the device in the erase state. The threshold voltage shift $\Delta V_{TH}$ was acquired by applying a linear fit to the linear regime of the reading $I_{DS}$-$V_{TG}$ curve. Figure 3a shows a clear shift of the threshold voltage when the width of the pulse is changed to 10 ms, which sets a reference for the following dynamic behavior measurements. $\Delta V_{TH}$ shows nearly a saturation behavior when the pulse width increases to 3 s (Figure 3a). The charge-trapping rate can be estimated from the expression $(dN_{trap}/dt) = (C_{HF-AL}/e)(\Delta V_{TH}/\Delta t)$, where $\Delta V_{TH}$ is the threshold voltage shift and $\Delta t$ is the pulse width. The calculated charge-trapping rate $(dV_G/dt)$ varies from 10$^{19}$ to 10$^{14}$ cm$^{-2}$ t$^{-1}$ when the pulse width changes from 1 ms to 3 s, which is much higher than other MoS$_2$-based memory devices[21]. The reason for such a high charge-trap rate is because of the thin Al$_2$O$_3$ tunnel layer (only 7 nm), which makes electron/hole charges much easier to tunnel through[44].

The retention characteristics of the device are determined by the height of the tunneling barrier and the depth of potential well formed in the Al$_2$O$_3$/HfO$_2$/Al$_2$O$_3$ charge-trap stack[45]. Figure 3c inset shows the threshold voltage at different time intervals after programming the device with a positive pulse (+26 V, duration of 3 s).

To prevent perturbation of the measurements to the device state, the transfer curve was also obtained in a small voltage range (-4 to +4 V). The extracted threshold voltage $\Delta V_{RTH}$ varies from 7.5 to 6.4 V after $10^4$ s (Figure 3c), from which we estimate that only ~28% of the charges will be lost after 10 years. The enhancement of the retention characteristics of our MoS$_2$ memory device is also related to the relatively clean interface between MoS$_2$ and the Al$_2$O$_3$ tunneling barrier.[14, 39] To test the endurance of the memory device, a sequence of pulse (±26 V, duration of 200 ms) was applied to the top gate with $V_{BG} = 0$ while $I_{DS}$ was measured ($V_{DS} = 50$ mV). As presented in Figure 3d, the device is stable after 120 cycles. The robustness and stability of the device shows a great perspective of applications in nonvolatile memory technology.

**Tunable characteristics of the MoS$_2$ memory device.** Memory cells with tunable characteristics can be used in multi-operational mode circuits where one device can show various functionalities.[46] To demonstrate such tenability, a back-gate voltage is applied to the MoS$_2$ channel and the transfer curves between -26 and +26 V were obtained under different $V_{BG}$ (Figure 4). It is noted that the window size can be effectively tuned from 15.6 to 21 V with $V_{BG}$ varying from +35 to -35 V (Figure 4 inset). The mechanism of the back-gate tunability can be qualitatively explained as follows: the negative $V_{BG}$ moves the Fermi level of the MoS$_2$ approaching the middle of the bandgap and makes it more intrinsic, thus becoming sensitive to the charges stored in the charge-trap layer, which also results in its being easily turned off. On the contrary, the positive $V_{BG}$ makes the MoS$_2$ channel highly doped and insensitive to the tunneling charges. The device is thus difficult to be switched off, leading to a smaller memory window.

To further investigate the tenability and the stability of the device under different back-gate voltage, the dynamic behavior was revisited. With $V_{DS} = 50$ mV, switching between the program and the erase state is achieved by applying negative and positive pulses to the top gate with different $V_{BG}$. As illustrated in Figure 5a, b and c, the device was set to a program state when a +26 V pulse with a duration of 200 ms was applied which makes the electrons accumulated in the charge-trap layer. We note that the charge density in the MoS$_2$ channel can suddenly increase during the positive pulse, as depicted by the source-drain current peak in Figure 5a, b and c. The device remains in the program state after the top-gate voltage is reset to 0 V; and the relatively low current corresponds to the device OFF state. After three seconds, a negative pulse (-26 V, 200 ms) is applied to set the device in the erase state. The high current reassembles the ON state. Importantly, there is no obvious decay or increase of the ON current, suggesting that the interface between MoS$_2$ and Al$_2$O$_3$ is clean[14, 39]. The application of $V_{BG}$ also leads to the change of the ON and OFF currents. When $V_{BG}$ =-35 V, the MoS$_2$ channel becomes more intrinsic with a relatively low conductance thus the ON and OFF current is low (Figure 5a). In comparison, When $V_{BG}$ =+35 V a high ON and OFF current level can be achieved (Figure 5c).

The retention of the program/erase state of the device was also measured with $V_{DS}$ = 50 mV under different $V_{BG}$. With a negative voltage pulse (-26 V, 3 s), the erase state current of the device was monitored in 2000 s. Subsequently, a positive voltage pulse (+26 V, 3 s) was applied to the top gate while the program current was recorded on the same time scale. Owing to the back-gate voltage, we can tune the level of program and erase current, for instance, $V_{BG}$ =-35 V enlarges the program/erase current ratio to the order of 10$^4$, making it easy for read-out; however, $V_{BG}$ =+35 V can dramatically reduce the current ratio to be about 3, demonstrating a wide range of tunability. Figure 5b, c and d also show six different current levels of the memory device and at least 2 bits can be stored in the memory device[47]. Compared to other multi-bit storage methods[48], the different current levels tuned by the back gate in the MoS$_2$ memory device are more controllable and stable which may allow for future applications of large-scale MoS$_2$-based nonvolatile memory devices.

To verify the hole storage capability in the charge-trap layer, the top-gate voltage was swept from -$V_{TG.max}$ to +$V_{TG.max}$ under a specific $V_{BG}$ of -35, 0, +35 V, as shown in Figure 5d, e and f, respectively. It is noted that the shift of the threshold voltage in the negative direction becomes smaller when $V_{BG}$ increases to +35 V. This is because the positive $V_{BG}$ electrostatically dopes the MoS$_2$ channel thus reducing the number of holes; so there are less holes tunneling though the barrier. The expansion of the memory window under $V_{BG}$ of -35 V is consistent with the results form Figure 4, evidencing a strong capability of the hole storage in our memory device.

In summary, we have demonstrated a dual-gate charge-trap memory device based on few-layer $MoS_2$ and $Al_2O_3/HfO_2/Al_2O_3$ charge-trap stack. Utilizing a back-gate structure, the memory window and the program/erase current ratio can be efficiently modulated. The robustness and stability of the tunable $MoS_2$-$HfO_2$ memory is promising for multi-bit information storage.

**Methods**

**Device fabrication and Electrical characterization.** Drain-source electrodes were fabricated by *e*-beam lithography (EBL) using PMMA/MMA bilayer polymer. Cr/Au (12nm/100nm) electrodes were deposited by *e*-beam evaporation. After lift-off, a 1 nm-thick Al layer was deposited and oxidized in the air for 24 hours, acting as a seed layer for subsequent deposition of $Al_2O_3/HfO_2/Al_2O_3$ (7nm/8nm/30nm) via atomic layer deposition (ALD). During the ALD process, trimethylaluminum and trtrakis (ethyl-methylamido) hafnium were reacted at 200 ℃ with water for $Al_2O_3$ and $HfO_2$, respectively. The top-gate (Cr/Au 12nm/100nm) electrode was fabricated using another EBL and metal deposition process. Electrical properties of the fabricated devices were measured in a probe station using a semiconductor device parameter analyzer (Agilent, B1500A).

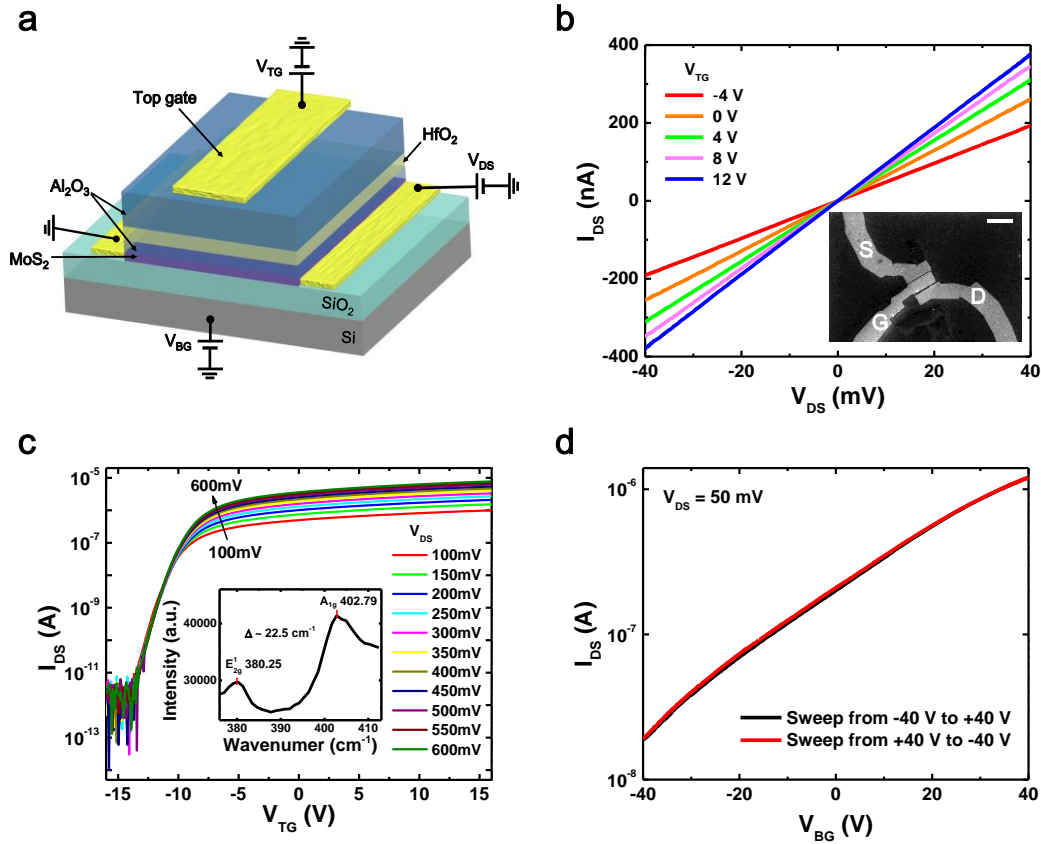

**Figure 1| A schematic structure of the MoS$_2$ charge-trap memory device and the field effect transistor (FET) performance. (a)** Dual-gate device structure. The few-layer MoS$_2$ and 8 nm-thick HfO$_2$ serve as the channel and the trap layer, respectively. **(b)** The output characteristics ($I_{DS}$-$V_{DS}$) of the device under different top-gate voltage. The inset shows a scanning electron microscopy (SEM) image of the memory device. Scale bar, 10 μm. **(c)** Transfer curves of the device with the top gate. The inset shows a Raman spectrum of the MoS$_2$ used in the device, which is determined to be 3-4 layers. **(d)** $I_{DS}$-$V_{BG}$ curves of the device with the back-gate voltage sweeping back and forth between -40 and +40 V, showing negligible interface charge states.

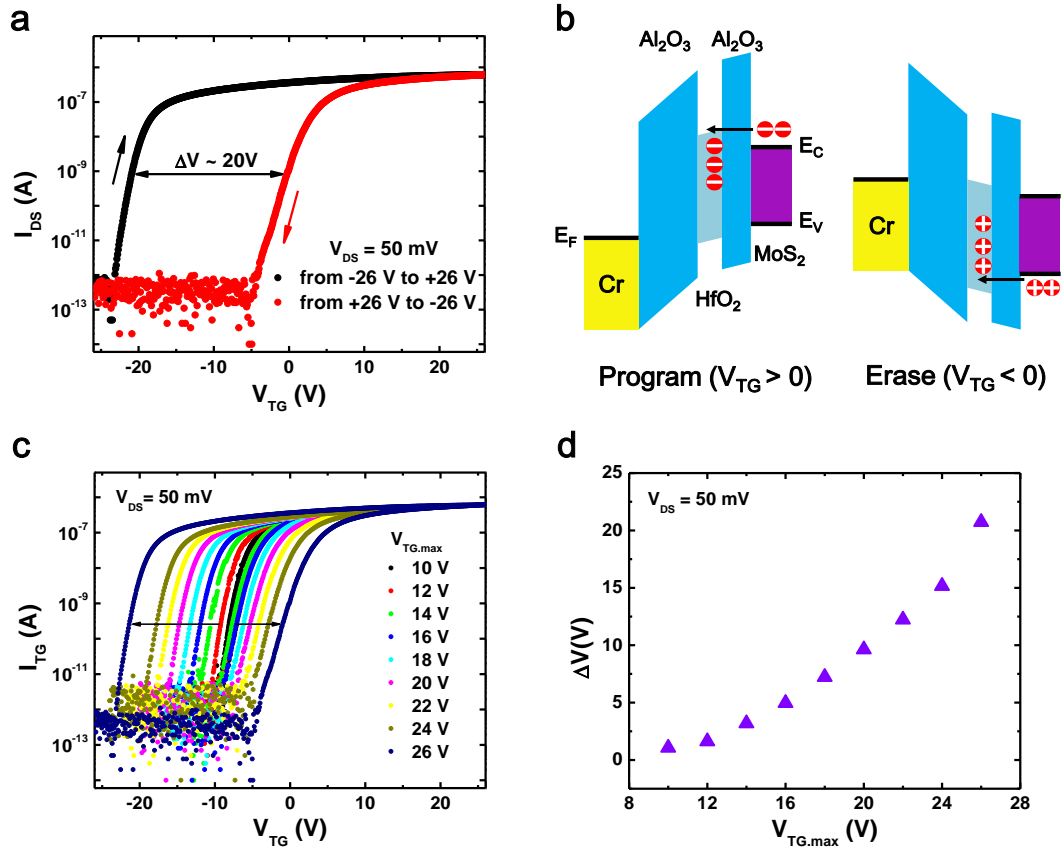

**Figure 2| Memory window and device band diagram. (a)** Threshold voltage shift when sweeping $V_{TG}$ from -26 to +26 V back and forth. The hysteresis of the transfer curves defines a memory window of ~20 V. **(b)** Band diagram of the program/erase state of the device under positive and negative $V_{TG}$. Positive $V_{TG}$ programs the device. Electrons tunneling from the few-layer $MoS_2$ channel are accumulated in the $HfO_2$ charge-trap layer. Negative $V_{TG}$ erases the device. Holes tunnel from the few-layer $MoS_2$ channel to the $HfO_2$ charge-trap layer. **(c)** $I_{TG}$ - $V_{TG}$ characteristics under different $V_{TG.Max}$ at $V_{DS}$ = 50 mV. **(d)** Extraction of memory window vs $V_{TG.Max}$. The memory window increases from ~1 to ~20 V in our experimental settings.

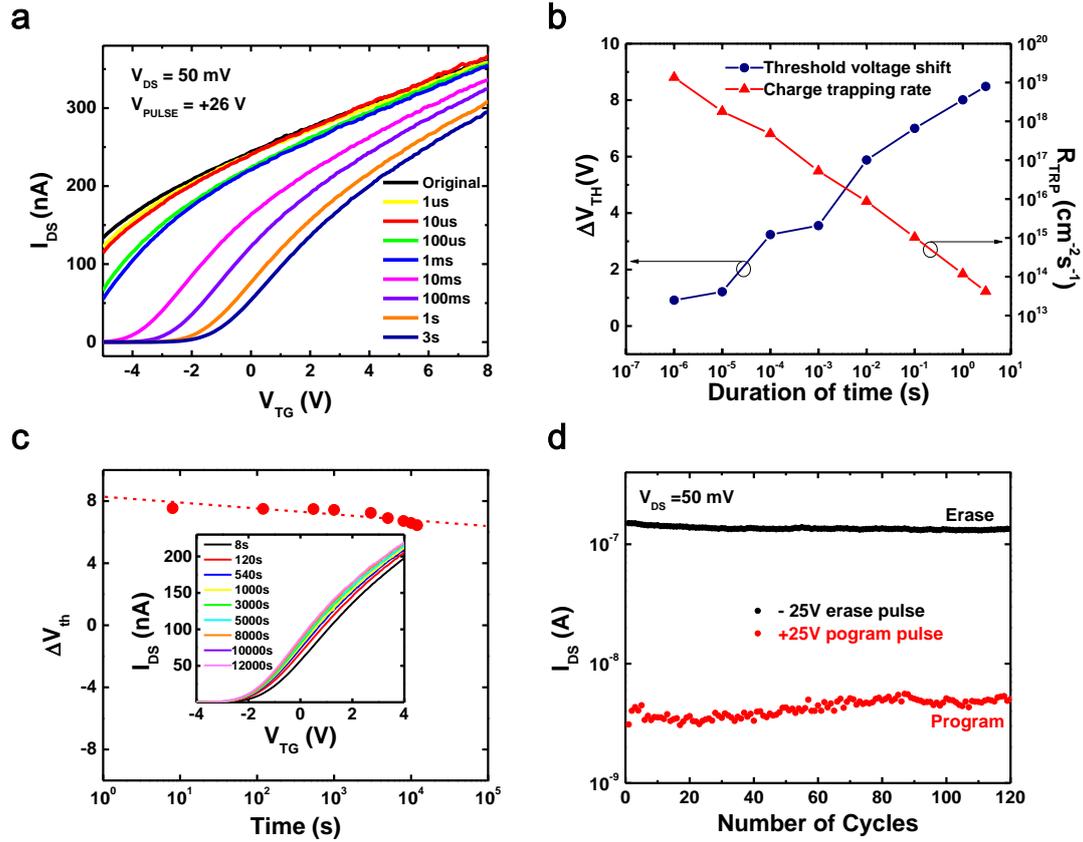

**Figure 3| Retention and endurance of the memory device. (a)** Transfer curves of $I_{DS}$-$V_{TG}$ in a narrow range of -5~+8 V under different pulse duration (1us-3s). **(b)** Extracted threshold voltage shift and calculated charge trap rate as a function of the pulse width. **(c)** Retention time of the threshold voltage. The programing pulse is set to be +26 V with 3 s duration. Inset shows the transfer curves ($I_{DS}$-$V_{TG}$) swept in different time intervals. Threshold voltage was obtained by linearly fitting to the transfer curves. We estimate that only 28% of the charges will be lost after 10 years. **(d)** Endurance of the memory device for 120 cycles with the program/erase voltage being +26 V, 200 ms duration and -26 V, 200 ms duration

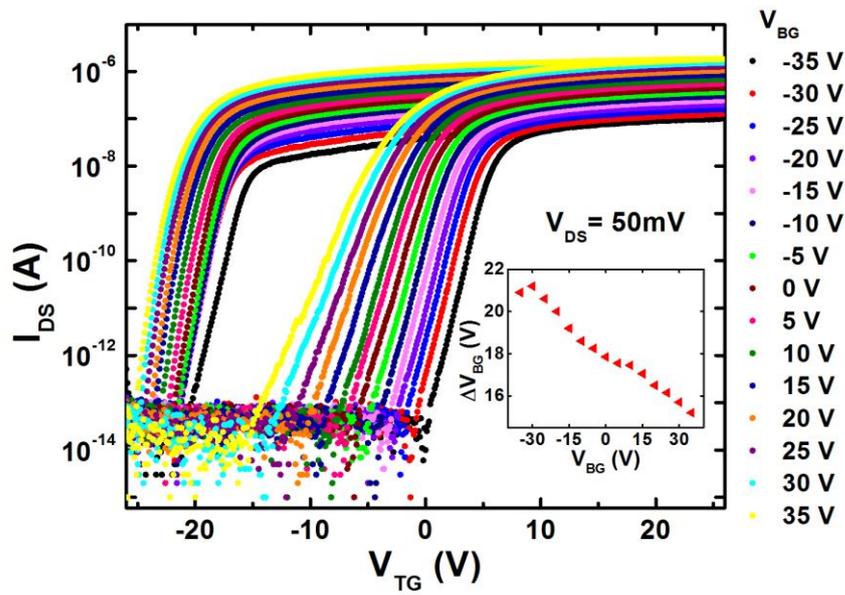

**Figure 4| Effect of back-gate voltage on the memory window.** The top gate of the device was swept from -26 to +26 V back and forth under different $V_{BG}$ ranging from -35 to +35 V. The inset summarizes the size of the memory window with respect to the $V_{BG}$.

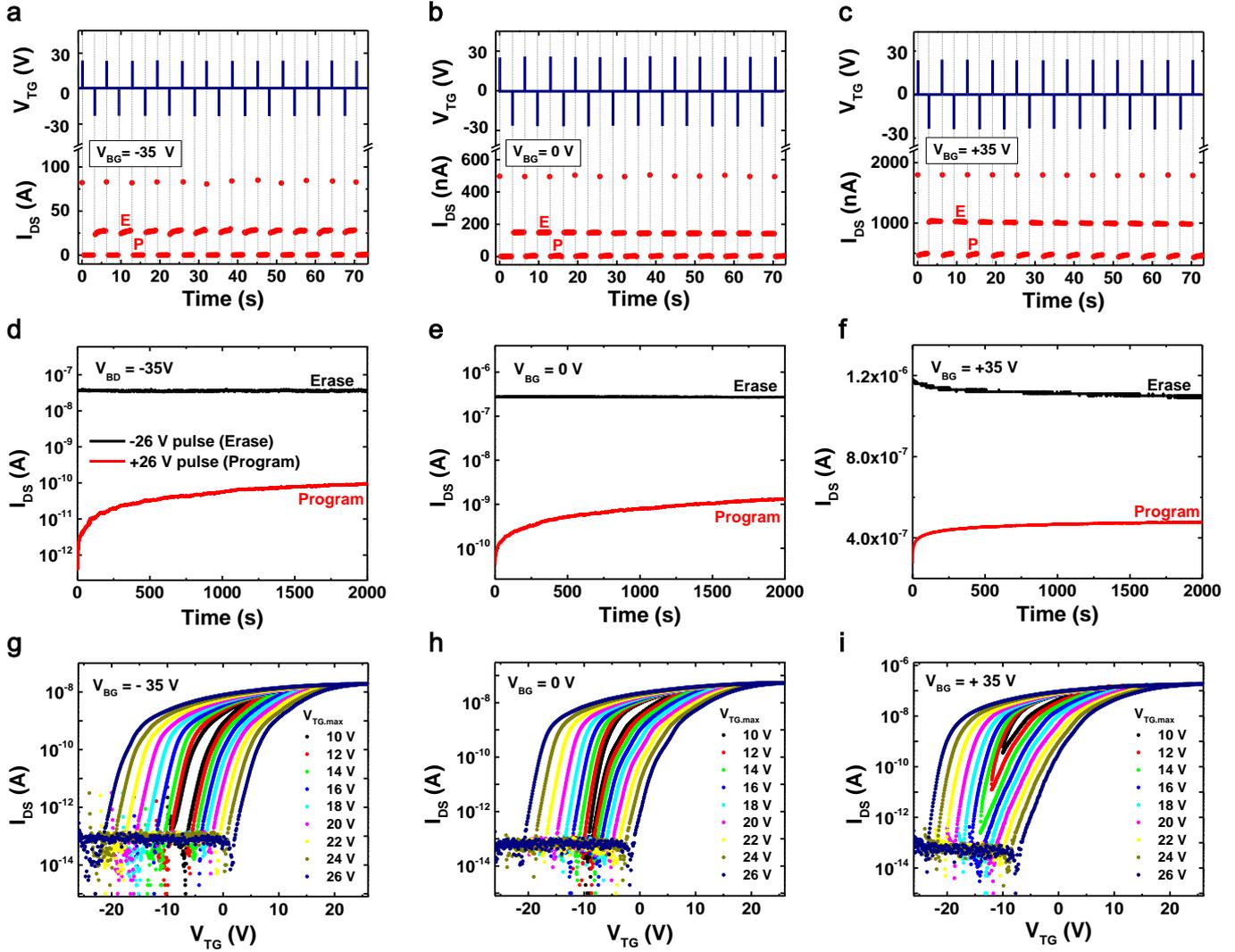

**Figure 5| Influence of the back-gate voltage on the memory characteristics. (a)**, (**b**) and **(c)**, the dynamic behavior of the device under a back-gate voltage of -35, 0 and +35 V, respectively. They were applied with a program voltage of +26 V ($V_{TG}$), 200 ms duration and an erase voltage of -26V, 200 ms duration ($V_{DS}$ = 50 mV). Negative back-gate voltage (-35 V) reduces the current levels of ON (Erase, device on) and OFF (Program, device off) states. In contrast, the positive back-gate voltage (+35 V) leads to higher ON and OFF current levels. **(d)**, **(e)** and **(f)**, the stability of program/erase state of the device after programing at +26 V, 3 s duration and erasing at -26 V, 3 s duration ($V_{DS}$ = 50 mV). The program/erase current ratio of the device varies from 3 to ~$10^4$ when the back-gate voltage changes from -35 to +35 V.**(g)**, **(h)** and **(i)**, significant change of the threshold voltage shift upon the application of the back-gate voltage ($V_{DS}$ = 30 mV). The negative part of the

threshold voltage shift corresponds to the hole trapping which can be suppressed by the positive back-gate voltage, vice versa for the electron trapping in the positive part.

26. Sharma SK, Prasad B, Kumar D. Application of high-k dielectric stacks charge trapping for CMOS technology. *Materials Science and Engineering: B* **166**, 170-173 (2010).

27. Hung M-F, Wu Y-C, Chang J-J, Chang-Liao K-S. Twin Thin-Film Transistor Nonvolatile Memory With an Indium–Gallium–Zinc–Oxide Floating Gate. *IEEE Electron Device Letters* **34**, 75-77 (2013).

28. Lo Y-S, Liu K-C, Wu J-Y, Hou C-H, Wu T-B. Bandgap engineering of tunnel oxide with multistacked layers of $Al_2O_3/HfO_2/SiO_2$ for Au-nanocrystal memory application. *Applied Physics Letters* **93**, 132907 (2008).

29. Reading MA, *et al.* High resolution medium energy ion scattering analysis for the quantitative depth profiling of ultrathin high-k layers. *Journal of Vacuum Science & Technology B: Microelectronics and Nanometer Structures* **28**, C1C65 (2010).

30. Chen W, Liu W-J, Zhang M, Ding S-J, Zhang DW, Li M-F. Multistacked $Al_2O_3/HfO_2/SiO_2$ tunnel layer for high-density nonvolatile memory application. *Applied Physics Letters* **91**, 022908 (2007).

31. Chang S, Song Y-W, Lee S, Lee SY, Ju B-K. Efficient suppression of charge trapping in ZnO-based transparent thin film transistors with novel $Al_2O_3/HfO_2/Al_2O_3$ structure. *Applied Physics Letters* **92**, 192104 (2008).

32. Suh DC, *et al.* Improved thermal stability of $Al_2O_3/HfO_2/Al_2O_3$ high-k gate dielectric stack on GaAs. *Applied Physics Letters* **96**, 142112 (2010).

33. Uk Lee D, Jun Lee H, Kyu Kim E, You H-W, Cho W-J. Low operation voltage and high thermal stability of a $WSi_2$ nanocrystal memory device using an $Al_2O_3/HfO_2/Al_2O_3$ tunnel layer. *Applied Physics Letters* **100**, 072901 (2012).

34. Jena D, Konar A. Enhancement of Carrier Mobility in Semiconductor Nanostructures by Dielectric Engineering. *Physical Review Letters* **98**,   (2007).

35. Chuang S, *et al.* $MoS_2$ P-type Transistors and Diodes Enabled by High Work Function $MoO_x$ Contacts. *Nano letters* **14**, 1337-1342 (2014).

36. Lee YT, *et al.* Graphene Versus Ohmic Metal as Source-Drain Electrode for $MoS_2$ Nanosheet Transistor Channel. *Small* **10**, 2356-2361 (2014).

37. Radisavljevic B, Kis A. Mobility engineering and a metal-insulator transition in monolayer $MoS_2$. *Nature materials* **12**, 815-820 (2013).

38. Li H, *et al.* From Bulk to Monolayer $MoS_2$: Evolution of Raman Scattering. *Advanced*
17